\documentclass{lmcs}
\pdfoutput=1
\usepackage[utf8]{inputenc}

\usepackage{lastpage}
\lmcsdoi{18}{3}{26}
\lmcsheading{}{\pageref{LastPage}}{}{}%
{Sep.~02,~2021}{Aug.~31,~2022}{}


\usepackage{algorithm}
\usepackage{booktabs}
\usepackage{algorithmic}
\algsetup{linenosize=\small}
\usepackage{amsfonts}
\usepackage{amsmath}
\usepackage{tabularx}
\usepackage[font=small,labelfont=bf]{caption}
\usepackage{capt-of}
\usepackage{float}
\usepackage{listings}
\usepackage{longtable}
\usepackage{mathtools}
\usepackage{marvosym}
\usepackage{newfloat}
\usepackage{subcaption}
\usepackage[inline]{enumitem}
\usepackage{url}
\usepackage{xcolor}
\usepackage{xspace}

\DeclareFloatingEnvironment[
  fileext = loe,
  listname = Examples,
  name = Example,
  placement = H,
  within = none,
  ]{customexample}

\definecolor{burgundy}{rgb}{.909,.125,.467}
\definecolor{dkgreen}{rgb}{0,0.6,0}
\definecolor{gray}{rgb}{0.5,0.5,0.5}
\definecolor{mauve}{rgb}{0.58,0,0.82}

\makeatletter
\RequirePackage[bookmarks,unicode,colorlinks=true]{hyperref}%
\def\@citecolor{blue}%
\def\@urlcolor{blue}%
\def\@linkcolor{blue}%

\def\orcidID#1{\smash{\href{http://orcid.org/#1}{\protect\raisebox{-1.25pt}{\protect\includegraphics{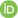}}}}}
\makeatother

\newboolean{showcomments}
\setboolean{showcomments}{true}

\newboolean{showidentity}
\setboolean{showidentity}{true}

\newboolean{longversion}
\setboolean{longversion}{true}

\newcommand{\SA}{\textbf{SA}\xspace}
\newcommand{\WA}{\textbf{WA}\xspace}
\newcommand{\Abs}[1]{\widehat{#1}\xspace}

\newcommand{\AbsRead}[2]{\Abs{\mathit{read}}(#1,#2)\xspace}
\newcommand{\AbsWrite}[3]{\Abs{\mathit{write}}(\Abs{#1},#2,#3)\xspace}

\newcommand{\constarr}[1]{\mathit{constarr(#1)}\xspace}
\newcommand{\Read}[2]{\mathit{read(#1,#2)}\xspace}
\newcommand{\Write}[3]{\mathit{write(#1,#2,#3)}\xspace}
\newcommand{\constarrnoarg}{\mathit{constarr}\xspace}
\newcommand{\Readnoarg}{\mathit{read}\xspace}
\newcommand{\Writenoarg}{\mathit{write}\xspace}
\newcommand{\arreq}{=\xspace}

\newcommand{\hist}[2]{\mathit{h_{#1}^{#2}}\xspace}

\newcommand{\widx}{\mathit{i_w}\xspace}
\newcommand{\ridx}{\mathit{i_r}\xspace}
\newcommand{\data}{\mathit{d}\xspace}
\newcommand{\ite}{\mathit{ite}\xspace}

\newcommand{\err}{\ensuremath{\mathit{err}}\xspace}

\newcommand{\ProphArray}[2]{\textbf{STS-CEGAR}(#1, #2)\xspace}

\newcommand{\Delay}[3]{\textbf{Delay}(#1, #2, #3)\xspace}
\newcommand{\Delaynoarg}{\textbf{Delay}\xspace}
\newcommand{\Prophecize}[4]{\textbf{Prophecize}(#1, #2, #3, #4)\xspace}
\newcommand{\Prophecizenoarg}{\textbf{Prophecize}\xspace}

\newcommand{\ArrayAxioms}[2]{\mathit{CheckArrayAxioms}(#1, #2)\xspace}
\newcommand{\ArrayAxiomsnoarg}{\textit{CheckArrayAxioms}\xspace}

\newcommand{\prophic}{\texttt{prophic3}\xspace}
\newcommand{\prophicSA}{\texttt{prophic3\_sa}\xspace}
\newcommand{\icthreeia}{\texttt{ic3ia}\xspace}
\newcommand{\freqhorn}{\texttt{freqhorn}\xspace}
\newcommand{\quicthree}{\texttt{quic3}\xspace}
\newcommand{\gspacer}{\texttt{gspacer}\xspace}
\newcommand{\Abstract}[1]{\textbf{Abstract}(#1)\xspace}
\newcommand{\Abstractnoarg}{\textbf{Abstract}\xspace}
\newcommand{\AbstractArrays}{\textbf{Abstract-Arrays}\xspace}

\newcommand{\RefineTS}[3]{\textbf{Refine}(#1, #2, #3)\xspace}
\newcommand{\RefineTSnoarg}{\textbf{Refine}\xspace}
\newcommand{\RefineTSArrays}{\textbf{Refine-Arrays}\xspace}
\newcommand{\Prove}[2]{\textbf{Prove}(#1, #2)\xspace}
\newcommand{\Provenoarg}{\textbf{Prove}\xspace}

\newcommand{\ca}{\mathit{ca}\xspace}
\newcommand{\nca}{\mathit{nca}\xspace}

\newcommand{\idxset}{\mathcal{I}\xspace}

\newcommand{\define}[1]{\textsl{#1}}
\newcommand{\Mo}{\mathbf{I}}
\newcommand{\Model}{\mathcal{M}}
\newcommand{\T}{\mathcal{T}}
\newcommand{\Prime}{\mathit{prime}}
\newcommand{\true}{\mathit{true}}
\newcommand{\ComputeIndices}{\mathit{ComputeIndices}}
\newcommand{\Int}{\mathit{Int}}
\newcommand{\Inv}{\mathit{Inv}}

\begin{document}

\title[Counterexample-Guided Prophecy]{Counterexample-Guided Prophecy for Model Checking Modulo the Theory of
  Arrays\rsuper*}
\titlecomment{{\lsuper*}This is an extended version of work presented in~\cite{cegp}.
}

\author[M.~Mann]{Makai Mann\lmcsorcid{0000-0002-1555-5784}}[a]
\author[A.~Irfan]{Ahmed Irfan\lmcsorcid{0000-0001-7791-9021}}[a]
\author[A.~Griggio]{Alberto Griggio\lmcsorcid{0000-0002-3311-0893}}[b]
\author[O.~Padon]{Oded Padon}[c]
\author[C.~Barrett]{Clark Barrett\lmcsorcid{0000-0002-9522-3084}}[a]

\address{Stanford University, Stanford, USA}
\email{\{makaim,irfan,barrett\}@cs.stanford.edu}
\address{Fondazione Bruno Kessler, Trento, Italy}
\email{griggio@fbk.eu}
\address{VMware Research, Palo Alto, USA}
\email{oded.padon@gmail.com}

\begin{abstract}
  We develop a framework for model checking infinite-state systems by
  automatically augmenting them with auxiliary variables, enabling
  quantifier-free induction proofs for systems that would otherwise require
  quantified invariants. We combine this mechanism with a counterexample-guided
  abstraction refinement scheme for the theory of arrays. Our framework can
  thus, in many cases, reduce inductive reasoning with quantifiers and arrays to
  quantifier-free and array-free reasoning. We evaluate the approach on a wide
  set of benchmarks from the literature. The results show that our
  implementation often outperforms state-of-the-art tools, demonstrating its
  practical potential.
\end{abstract}

\maketitle

\section{Introduction}

Model checking is a widely-used and highly-effective technique for automated property checking.
While model checking finite-state systems is a
well-established technique for hardware and software systems, model checking infinite-state
systems is more challenging.  One challenge, for example, is that proving
properties by induction over infinite-state systems often requires the use of
universally quantified invariants.
While some automated reasoning tools can reason 
about quantified
formulas, such reasoning is typically not very robust.  Furthermore, just
discovering these quantified invariants remains very challenging.

Previous work (e.g.,~\cite{eager-abstraction}) has shown that prophecy variables
can sometimes play the same role as universally quantified variables, making it
possible to transform a system that would require quantified reasoning into one
that does not. However, to the best of our knowledge, there has been no
automatic method for applying such transformations. In this paper, we introduce
a technique we call \emph{counterexample-guided prophecy}. During the refinement
step of an abstraction-refinement loop, our technique automatically introduces
prophecy variables, which both help with the refinement step and may also reduce
the need for quantified reasoning. We demonstrate the technique in the context
of model checking for infinite-state systems with arrays, a domain which is
known for requiring quantified reasoning. We show how a standard abstraction for
arrays can be augmented with counterexample-guided prophecy to obtain an
algorithm that reduces the model checking problem to quantifier-free, array-free
reasoning. The paper makes the following contributions:
\begin{enumerate}[i)]
\item we introduce an algorithm called \Prophecizenoarg that uses history and
  prophecy variables to target a specific term at a specific time step of an
  execution, producing a new transition system that can effectively reason
  universally about that term;
\item we develop an automatic abstraction-refinement procedure for arrays,
  which leverages the \Prophecizenoarg algorithm during the refinement step, and
  show that it is sound and produces no false positives;
\item we develop a prototype implementation of our technique; and
\item we evaluate our technique on four sets of model checking benchmarks
  containing arrays and show that our implementation outperforms
  state-of-the-art tools on a majority of the benchmark sets.
\end{enumerate}

\iflongversion
The rest of the paper is organized as follows. We start by providing relevant
background information in Section~\ref{sec:background}. We then motivate the use
of prophecy variables with an example and introduce the $\Prophecizenoarg$
algorithm in Section~\ref{sec:motivation}. We describe our
abstraction-refinement framework for arrays in Section~\ref{sec:cegarloop} and
discuss its expressiveness and limitations in Section~\ref{sec:expressiveness}.
In Section~\ref{sec:prototype}, we describe our prototype along with some
implementation details.
We evaluate our approach empirically in Section~\ref{sec:experiments}, cover
related work in Section~\ref{sec:related}, and finally conclude in
Section~\ref{sec:conclusion}.

This paper is an extended version of work presented in~\cite{cegp}. Notable
changes include the proofs, a self-comparison with different options in
Section~\ref{sec:experiments}, and Section~\ref{sec:prototype}, which discusses
implementation details of the prototype.

\fi

\section{Background}
\label{sec:background}

We assume the standard many-sorted first-order logical setting with the usual
notions of signature, term, formula, and interpretation.
A \define{theory} is a pair $\T = (\Sigma, \Mo)$ where
$\Sigma$ is a signature and  $\Mo$ is a class of $\Sigma$-interpretations,
the \define{models} of $\T$.
A $\Sigma$-formula $\varphi$ is
\define{satisfiable} (resp., \define{unsatisfiable}) \define{in $\T$}
if it is satisfied by some (resp., no) interpretation in $\Mo$.
Given an interpretation $\Model$, a variable assignment $s$ over a set of
variables $X$ is a mapping that assigns each variable $x\in X$ of sort $\sigma$
to an element of $\sigma^{\Model}$, denoted $x^s$.  We write $\Model[s]$ for the interpretation
that is equivalent to $\Model$ except that each variable $x\in X$ is mapped to $x^s$.
Let $x$ be a variable,
$t$ a term, and $\phi$ a formula. We denote with $\phi\{x \mapsto t\}$ the
formula obtained by replacing every free occurrence of $x$ in $\phi$ with $t$. We
extend this notation to sets of variables and terms in the usual way.  If $f$
and $g$ are two functions, we write $f \circ g$ to mean functional composition, i.e.,
$f \circ g (x) = f(g(x))$. We use $\uplus$ to refer to set union.

Let $\T_A$ be the standard theory of arrays~\cite{mccarthy} with extensionality,
extended with constant arrays. Concretely, we assume sorts for arrays,
indices, and elements, and function symbols $\Readnoarg$, $\Writenoarg$, and
$\constarrnoarg$. Here and below, we use $a$ and $b$ to refer to arrays, $i$ and
$j$ to refer to array indices, and $e$ and $c$ to refer to array elements, where
$c$ is also restricted to be an interpreted constant. The theory contains the
class of all interpretations satisfying the following axioms:

\begin{equation}
  \tag{write}
  \label{eq:array-write}
  \begin{aligned}
    \forall\, a, i, j, e.\: & i = j \implies \Read{\Write{a}{j}{e}}{i} = e ~\wedge \\
    & i \neq j \implies \Read{\Write{a}{j}{e}}{i} = \Read{a}{i}
  \end{aligned}
\end{equation}
\begin{equation}
  \tag{ext}
  \label{eq:array-ext}
    \forall\, a, b.\: (\forall\, i.\: \Read{a}{i} = \Read{b}{i}) \implies a \arreq b
\end{equation}
\begin{equation}
  \tag{const}
  \label{eq:array-const}
  \forall\, i.\: \Read{\constarr{c}}{i} = c
\end{equation}

\subsection{Symbolic Transition Systems and Model Checking}
For generality, assume
a background theory $\T$ with signature $\Sigma$. We will assume that all terms
and formulas are $\Sigma$-terms and $\Sigma$-formulas, that entailment is
entailment modulo $\T$, and interpretations are $\T$-interpretations. A symbolic
transition system (STS) $\mathcal{S}$ is a tuple $\mathcal{S} := \langle X, I, T
\rangle $, where $X$ is a finite set of state variables, $I(X)$ is a formula
denoting the initial states of the system, and $T(X, X')$ is a formula
expressing a transition relation. Here, $X'$ is the set obtained by replacing
each variable $x\in X$ with a new variable $x'$ of the same sort. Let $\Prime(x)
= x'$ be the bijection corresponding to this replacement. We say that a variable
$x$ is \textit{frozen} if $T \models x' = x$. When the state variables are
obvious, we will often drop $X$.

A state $s$ of $\mathcal{S}$ is a variable assignment over $X$. An
\emph{execution} of $\mathcal{S}$ of length $k$ is a pair $\langle
\Model,\pi\rangle$, where $\Model$ is an interpretation and $\pi := s_0, s_1,
\ldots, s_{k-1}$ is a \emph{path} of length $k$, a sequence of states such that
$\Model[s_0] \models I(X)$ and $\Model[s_i][s_{i+1} \circ \Prime^{-1}] \models
T(X, X')$ for all $0 \leq i < k - 1$. When reasoning about paths, it is often
convenient to have multiple copies of the state variables $X$. We use $X@n$ to
denote the set of variables obtained by replacing each variable $x\in X$ with a
new variable called $x@n$ of the same sort. We refer to these as \emph{timed}
variables. A state $s$ is \emph{reachable} in $\mathcal{S}$ if it appears in a
path of some execution of $\mathcal{S}$. We say that a formula $P(X)$ is an
\emph{invariant} of $\mathcal{S}$, denoted by $\mathcal{S} \models P(X)$, if
$P(X)$ is satisfied in every reachable state of $\mathcal{S}$ (i.e., for every
execution $\langle \Model,\pi \rangle$, $\Model[s] \models P(X)$ for each $s$ in
$\pi$). The \emph{invariant checking problem} is, given $\mathcal{S}$ and
$P(X)$, to determine if $\mathcal{S} \models P(X)$. A \emph{counterexample} is
an execution $\langle \Model,\pi \rangle$ of $\mathcal{S}$ of length $k$ such
that $\Model[s_{k-1}] \not\models P(X)$. If $I(X) \models \phi(X)$ and $\phi(X) \wedge
T(X, X') \models \phi(X')$, then $\phi(X)$ is an \emph{inductive} invariant. Every
inductive invariant is an invariant (by induction over path length). In this
paper we focus on model checking problems where $I$, $T$ and $P$ are quantifier-free.
However, a \emph{quantified inductive invariant} might still be necessary to
prove a property of the system.

\emph{Bounded Model Checking} (BMC) is a bug-finding technique which attempts to
find a counterexample for a property, $P(X)$, of length $k$ for some finite
$k$~\cite{bmc}. A single BMC query at bound $k$ for an invariant property uses a
constraint solver to check the satisfiability of the following formula: $
BMC(\mathcal{S}, P, k) \coloneqq I(X@0) \wedge (\bigwedge_{i=0}^{k-2} T(X@i,
X@(i+1))) \wedge \neg P(X@(k-1))$. If the query is satisfiable, there is a bug.

\subsection{Counterexample-Guided Abstraction Refinement (CEGAR)} CEGAR
is a general technique in which a difficult
conjecture is tackled iteratively~\cite{cegar}. Algorithm~\ref{alg:proph-array}
shows a simple CEGAR loop for checking an invariant $P$ for an STS
$\mathcal{S}$. It is parameterized by three functions. The \Abstractnoarg
function produces an initial abstraction of the problem. It must satisfy the
contract that if $\langle \Abs{S}, \Abs{P}\rangle = \Abstract{\mathcal{S},P}$,
then $\Abs{S} \models \Abs{P} \implies \mathcal{S} \models P$. The next function is the
\Provenoarg function. This can be any (unbounded) model-checking algorithm that
can return counterexamples. It checks whether a given property $P$ is an
invariant of a given STS $\mathcal{S}$. If it is, it returns with
$\mathit{proven}$ set to true. Otherwise, it returns a bound $k$ at which a
counterexample exists. The final function is \RefineTSnoarg. It takes the
abstracted STS and property together with a bound $k$ at which a known
counterexample for the abstract STS exists. Its job is to refine the abstraction
until there is no longer a counterexample of size $k$. If it succeeds, it
returns the new STS and property. It fails if there is an actual counterexample
of size $k$ for the concrete system. In this case, it sets the return value
$\mathit{refined}$ to false.

\begin{algorithm}[t]
  \caption{\ProphArray{$\mathcal{S} \coloneqq \langle X, I, T \rangle$}{$P$}}
  \label{alg:proph-array}
  \begin{algorithmic}[1]
    \STATE $\langle \langle \Abs{X}, \Abs{I}, \Abs{T} \rangle, \Abs{P}\rangle \gets
    \Abstract{\mathcal{S},P}$
    \WHILE{true}
    \STATE $\langle k, \mathit{proven} \rangle \gets
    \Prove{\langle \Abs{X}, \Abs{I}, \Abs{T}\rangle}{\Abs{P}}$ \COMMENT{try to prove}

    \STATE \textbf{if} \textit{proven} \textbf{then return} true \COMMENT{property proved}
    \STATE $\langle \langle \Abs{X}, \Abs{I}, \Abs{T} \rangle,
    \Abs{P}, \mathit{refined} \rangle \gets
    \RefineTS{\langle \Abs{X}, \Abs{I}, \Abs{T} \rangle}{\Abs{P}}{k}$ \COMMENT{try
      to refine}
	  \STATE \textbf{if} $\neg\textit{refined}$ \textbf{then return} false \COMMENT{found counterexample}
    \ENDWHILE
  \end{algorithmic}
\end{algorithm}

\subsection{Auxiliary variables}
\label{sec:auxiliaryvars}
We finish this section with relevant background on \emph{auxiliary} variables, a
crucial part of the refinement step described in Section~\ref{sec:cegarloop}.
Auxiliary variables are new variables added to the system which do not influence
its behavior (i.e., a state is reachable in the old system iff it is a
reduct~\cite{model-theory} to the old set of variables of a reachable state in
the new system). There are two main categories of auxiliary variables we
consider: \textit{history} and \textit{prophecy}. History variables, also known
as \textit{ghost state}, preserve a value, making its past value available in
future states~\cite{hist-vars}. Prophecy variables are the dual of history
variables and provide a way to refer to a value that occurs in a future state.
Abadi and Lamport formally characterized soundness conditions for the
introduction of history and prophecy variables~\cite{abadi-lamport}. Here, we
consider a simple, structured form of history variables.
\begin{defi}
  \label{def:delay}
  Let $\mathcal{S} = \langle X, I, T \rangle$ be an STS, $t$ a term whose free
  variables are in $X$, and $n > 0$, then $\Delay{\mathcal{S}}{t}{n}$ returns a
  tuple $\langle \langle X^h, I^h, T^h \rangle, \hist{t}{n} \rangle$, containing
  a new STS and history variable, where $X^h = X \uplus
  \{\hist{t}{1},\ldots,\hist{t}{n}\}$, $I^h = I$, and $T^h = T \wedge
  \hist{t}{1}' = t \wedge \bigwedge_{i=2}^{n} \hist{t}{i}' = \hist{t}{i-1}$.
\end{defi}
The $\Delaynoarg$ operator makes the current value of a term $t$ available for
the next $n$ states in a path.  This is accomplished by adding $n$ new history
variables and creating an assignment chain that passes the value to the next
history variable at each state.  Thus, $\hist{t}{k}$ contains the
value that $t$ had $k$ states ago. The initial value of each history variable is
unconstrained.

\begin{thmC}[\cite{abadi-lamport}]
\label{thm:hist}
Let $\mathcal{S} = \langle X, I, T \rangle$ be an STS, $P$ a property, and
$\Delay{\mathcal{S}}{v}{n} = \langle \mathcal{S}^h, \hist{v}{n}\rangle$. Then $\mathcal{S} \models
P$ \textit{iff} $\mathcal{S}^h \models P$.
\end{thmC}

\noindent
We refer to~\cite{abadi-lamport} for a general proof that subsumes Theorem~\ref{thm:hist}.
In contrast to the general approach for history variables, we
use a version of prophecy that only requires a single frozen variable. The
motivation for this is that a frozen variable can be used in place of a
universal quantifier, as the following theorem adapted
from~\cite{eager-abstraction} shows.

\begin{thmC}[\cite{eager-abstraction}]
  \label{thm:prophecy}
  Let $\mathcal{S} = \langle X, I, T \rangle$ be an STS, $y$ a variable in
  formula $P(X \cup \{y\})$, and $v$ a fresh variable (i.e., not in $X \cup \{y\}$
  or $X'$). Let $\mathcal{S}^p = \langle X\cup\{v\}, I, T \wedge v'=v
  \rangle$. Then $\mathcal{S} \models \forall\,y.\: P(X \cup \{y\})$ \textit{iff}
  $\mathcal{S}^p \models P(X \cup \{y\})\{y \mapsto v\}$.
\end{thmC}

\iflongversion
\noindent\textit{Proof Sketch.}
We prove the equivalent statement $\mathcal{S} \not\models \forall\,y.\: P(X
\cup \{y\}) \leftrightarrow \mathcal{S}^p \not\models P(X \cup \{y\})\{y \mapsto
v\}$.
\\
\noindent$\rightarrow$: Suppose there is a counterexample trace demonstrating that $\mathcal{S}
\not\models \forall\,y.\: P(X \cup \{y\})$. In the last step of the trace, the
property is violated. There is a specific value of $y$ for which $P(X \cup
\{y\})$ does not hold. We can reconstruct the same counterexample trace for
$\mathcal{S}^p$ and $P(X \cup \{y\})\{y \mapsto v\}$ by assigning $v$ this same
value in every step of the trace.
\\
\noindent$\leftarrow$: If there is a counterexample trace for $\mathcal{S}^p
\not\models P(X \cup \{y\})\{y \mapsto v\}$, we can construct a counterexample
for $\mathcal{S}$ and $\forall\,y.\: P(X \cup \{y\})$ by using the same trace
but dropping the assignment to the frozen variable, $v$. Finally, the value of
$y$ that violates the property in the last step of the trace will be the same
value $v$ was assigned. \hfill$\Box$
\fi
\\\\
\noindent
Theorem~\ref{thm:prophecy} shows that a universally quantified variable in a
property can be replaced with a fresh symbol in a process similar to
Skolemization. The intuition is as follows. The frozen variable has the same
value in all states, but it is uninitialized by $I$. Thus, for each path in
$\mathcal{S}$, there is a corresponding path (i.e., identical except at $v$) in
$\mathcal{S}^p$ for \emph{every} possible value of $v$. This proliferation of
paths plays the same role as the quantified variable in $P$. We mention here one
more theorem from~\cite{eager-abstraction}. This one allows us to
\emph{introduce} a universal quantifier.

\begin{thmC}[\cite{eager-abstraction}]
  \label{thm:intro-univ}
  Let $\mathcal{S} = \langle X, I, T \rangle$ be an STS, $P(X)$
  a formula, and $t$ a term. Then, $\mathcal{S} \models P(X)$ \textit{iff}
  $\mathcal{S} \models \forall\,y. (y = t \implies P(X))$, where $y$ is not
  free in $P(X)$.
\end{thmC}
\iflongversion
\noindent\textit{Proof Sketch.}
$P(X)$ and $\forall\,y. (y = t \implies P(X))$ for $y \not\in X$ are equivalent
in first order logic. Intuitively, when $y = t$, it simplifies to $P(X)$ and all
other values of $y$ render the formula trivially true. \hfill$\Box$
\fi
\\\\
\noindent
Theorems~\ref{thm:prophecy} and~\ref{thm:intro-univ} are special cases of
Theorems 3 and 4 of~\cite{eager-abstraction}, which handle temporal
logic~\cite{ltl} formulas. Another notable difference is that Theorem 3
of~\cite{eager-abstraction} uses a fresh background symbol to replace the
universally quantified variable. Note that it does not change as the system
evolves because it is not a state variable of the transition system. Rather than
allowing background symbols, we simulate this with a frozen variable that
maintains its value in Theorem~\ref{thm:prophecy}.

\section{Using Auxiliary Variables to Assist Induction}
\label{sec:motivation}

We can use Theorem~\ref{thm:intro-univ} followed by Theorem~\ref{thm:prophecy}
to introduce frozen prophecy variables that predict the value of a term $t$ when
the property $P$ is being checked. We refer to $t$ as the prophecy
\textit{target} and the process as \textit{universal} prophecy. If we also use
$\Delaynoarg$, we can target a term at some finite number of steps
\emph{before} the property is checked. This is captured by
Algorithm~\ref{alg:prophecize}, which takes a transition system, property $P(X)$, term
$t$, and $n \geq 0$. If $n = 0$, it introduces a universal prophecy variable for
$t$. Otherwise, it first introduces history variables for $t$ and then
applies universal prophecy to the delayed $t$. In either case it returns the
augmented system, augmented property, and the prophecy variable.

\begin{algorithm}[t]
  \caption{$\Prophecize{\langle X, I, T \rangle}{P(X)}{t}{n}$}
  \label{alg:prophecize}
  \begin{algorithmic}[1]
    \IF{n = 0}
    \RETURN $\langle \langle X \uplus \{p_t\}, I, T \wedge p_t' = p_t \rangle, p_t
    = t \implies P(X), p_t \rangle$
    \ELSE
    \STATE $\langle \langle X^h, I^h, T^h \rangle, \hist{t}{n} \rangle \coloneqq \Delay{\langle X, I, T \rangle}{t}{n}$
    \RETURN $\langle \langle X^h \uplus \{p_t^n\}, I, T \wedge p_t^{n'} = p_t^n \rangle, p_t
    = \hist{t}{n} \implies P(X), p_t^n \rangle$
    \ENDIF

  \end{algorithmic}
\end{algorithm}

We will use the STS shown in Fig.~\ref{fig:running}(a) as a running example
throughout the paper (it is inspired by the hardware example
from~\cite{bjesse}). We assume the background theory $\T$ includes integer
arithmetic and arrays of integers indexed by integers. The variables in this STS
include an array and four integer variables, representing the read index ($\ridx$), write
index ($\widx$), read data ($\data_r$), and write data ($\data_w$), respectively. The system starts with an array
of all zeros. At every step, if the write data is less than 200, it writes that
data to the array at the write index. Otherwise, the array stays the same.
Additionally, the read data is updated with the current value of $a$ at $\ridx$.
This effectively introduces a one-step delay between when the value is read from
$a$ and when the value is present in $\data_r$. The property is that $\data_r <
200$. This property is clearly true, but it is not straightforward to prove with
standard model checking techniques because it is not inductive. Note that it is
also not $k$-inductive for any~$k$~\cite{k-induction}. The primary issue is that
the property does not constrain the value of $a$ at all, so in an inductive proof, the
value of $a$ could be anything in the induction hypothesis.

\begin{figure*}[t]
	\subcaptionbox{}{
		\begin{minipage}{.45\textwidth}
			\small
			\begin{equation*}
			\begin{aligned}
			I \coloneqq \ & a \arreq \constarr{0} \wedge \data_r < 200  \\
			T \coloneqq \ &  a' \arreq \ite(\data_w < 200, \\
			& \hspace{4em} \Write{a}{\widx}{\data_w}, a) \wedge \\
			& \data_r' = \Read{a}{\ridx} \\
			P \coloneqq \ & \data_r < 200
			\end{aligned}
			\end{equation*}
		\end{minipage}
	}
	\subcaptionbox{}{
		\begin{minipage}{.45\textwidth}
			\small
			\begin{equation*}
			\begin{aligned}
			I \coloneqq \ & a \arreq \constarr{0} \wedge \data_r < 200  \\
			T \coloneqq \ & a' \arreq \ite(\data_w < 200, \\
			& \hspace{4em} \Write{a}{\widx}{\data_w}, a) \wedge \\
			\data_r' = & \Read{a}{\ridx} \wedge p_{\ridx}^{1'} = p_{\ridx}^{1} \wedge \hist{\ridx}{1}' = \ridx \\
			P \coloneqq \ & p_{\ridx}^{1} = \hist{\ridx}{1} \implies \data_r < 200
			\end{aligned}
			\end{equation*}
		\end{minipage}
	}
	\caption{(a) Running example. (b) Running example with prophecy
    variable. \label{fig:running}}
\end{figure*}

One way to prove the property is to strengthen it with the quantified invariant:
$\forall\, i.\: \Read{a}{i} < 200$. Remarkably, observe that by augmenting the
system using \Prophecizenoarg, it is possible to prove the property using only a
\emph{quantifier-free} invariant. In this case, the relevant prophecy target is
the value of $\ridx$ one step before checking the property. We run
$\Prophecize{\langle X, I, T \rangle}{P}{i_r}{1}$ and it returns the system and
property shown in Fig.~\ref{fig:running}(b), along with the prophecy variable
$p_{\ridx}^1$. This augmented system has a simple, quantifier-free invariant
which can be used to strengthen the property, making it inductive:
$\Read{a}{p_\ridx} < 200$. This formula holds in the initial state because of
the constant array, and if we start in a state where it holds, it still holds
after a transition.

Notice that the invariant learned over the prophecy variable has the same form
as the original quantified invariant. However, we have instantiated that
universal quantifier with a fresh, frozen prophecy variable. Intuitively, the
prophecy variable captures a proof by contradiction: assume the property does
not hold, consider the value of $\ridx$ one step before the first failure of the
property, and then use this value to show the property holds. This example shows
that auxiliary variables can be used to transform an STS without a
quantifier-free inductive invariant into an STS with one. However, it is not yet
clear how to identify good targets for history and prophecy variables. In the
next section, we show how this can be done as part of an abstraction refinement
scheme for symbolic transition systems over the theory of arrays.

\section{Abstraction Refinement for Arrays}
\label{sec:cegarloop}

We now introduce our main contribution.  Given a background
theory $\T_B$ and a model checking algorithm for STSs over $\T_B$, we use
an instantiation of the CEGAR loop in Algorithm~\ref{alg:proph-array} to
check properties of STSs over the theory that combines
$\T_B$ and the theory of arrays, $\T_A$.  The key idea is to abstract all array
operators and then add array lemmas as needed during refinement.

\subsection{Abstract and Prove}
We use a standard abstraction for the theory of
arrays, which we denote $\AbstractArrays$. Every array sort is replaced with an
uninterpreted sort, and the array variables are abstracted accordingly. Each
constant array is replaced by a fresh abstract array variable, which is then
constrained to be frozen (because constant arrays do not change over time).
Additionally, we replace the $\Readnoarg$ and $\Writenoarg$ array operations
with uninterpreted functions. Note that if the system contains multiple array
sorts, we need to introduce separate read and write functions for each
uninterpreted abstract array sort. Using uninterpreted sorts and functions for
abstracting arrays is a common technique in Satisfiability Modulo
Theories~\cite{smt} (SMT) solvers~\cite{array-form}. Intuitively, our initial
abstraction starts with \textit{memoryless} arrays, i.e., the array axioms are
not initially enforced on the abstraction. We then incrementally refine the
arrays' memory as needed by adding prophecy variables to be used in array
axioms. Intuitively, a prophecy variable keeps track of an index of the array
that will faithfully store values. Fig.~\ref{fig:abs-running} shows the result
of running $\AbstractArrays$ on the example from Fig.~\ref{fig:running}(a).
\Provenoarg can be instantiated with any (unbounded) model checker that can
accept expressions over the background theory $\T_B$ combined with the theory of
uninterpreted functions. In particular, due to our abstraction, the model
checker does not need to support the theory of arrays.

\begin{figure}[t]
  \small
  \begin{equation*}
    \begin{aligned}
      \Abs{I} \coloneqq \ & \Abs{a} \arreq \Abs{\mathit{constarr0}} \wedge \data_r < 200 \\
      \Abs{T} \coloneqq \ & \Abs{a}' \arreq ite(\data_w < 200, \AbsWrite{a}{\widx}{\data_w}, \Abs{a}) \wedge \\
      & \data_r' = \AbsRead{\Abs{a}}{\ridx} \wedge \Abs{\mathit{constarr0}}' = \Abs{\mathit{constarr0}} \\
      \Abs{P} \coloneqq \ & \data_r < 200
    \end{aligned}
  \end{equation*}
  \caption{Result of calling \Abstractnoarg on the example from Fig.~\ref{fig:running}(a) \label{fig:abs-running}}
  \end{figure}

\subsection{Refine}
Here, we explain the refinement approach for our array
abstraction. At a high level, the algorithm solves a BMC problem over the
abstract STS at bound $k$. It identifies violations of array axioms
in the returned abstract counterexample, and instantiates each violated axiom (this is
essentially the same as the lazy array axiom instantiation approach used in SMT
solvers~\cite{bradley-arrays,lemmas-on-demand,weakly-eq-arrays,general-arrays}).
We then \emph{lift} these axioms to the STS-level by modifying the STS.\@ It is
this step that may require introducing auxiliary variables. The details are
shown in Algorithm~\ref{alg:refine-ts}.

\begin{algorithm}[t]
  \caption{\RefineTSArrays($\Abs{\mathcal{S}} \coloneqq \langle \Abs{X}, \Abs{I},
      \Abs{T} \rangle, \Abs{P}, k$)}
  \label{alg:refine-ts}
  \begin{algorithmic}[1]
    \STATE $\idxset \gets \ComputeIndices(\Abs{\mathcal{S}},\Abs{P},k)$
    \LOOP
    \STATE{$\rho \gets$ BMC($\Abs{\mathcal{S}},\Abs{P},k$)}

    \STATE \textbf{if} $\rho = \bot$ \textbf{then return} $\langle \langle \Abs{X}, \Abs{I}, \Abs{T} \rangle, \Abs{P}, \true \rangle$ \COMMENT{Property holds up to bound $k$}
    \STATE $\langle \ca, \nca \rangle \gets \ArrayAxioms{\rho}{\idxset}$

    \STATE \textbf{if} $\ca = \emptyset \wedge \nca = \emptyset$ \textbf{then return} $\langle \langle \Abs{X}, \Abs{I}, \Abs{T} \rangle, \Abs{P}, false \rangle$ \COMMENT{True counterexample}
    \STATE // Go through non-consecutive array axiom instantiations
    \FOR{$\langle ax, i@n_i\rangle \in \nca$}
    \STATE \textbf{let} $n_{min} \coloneqq \mathit{min}(\tau(ax) \backslash \{n_i\})$
    \STATE $\langle \langle X^p, I^p, T^p \rangle, P^p, p_i^{k-n_i} \rangle \gets
    \Prophecize{\langle \Abs{X}, \Abs{I}, \Abs{T} \rangle}{\Abs{P}}{i}{k-n_i}$
    \STATE $ax_c \gets ax\{i@n_i \mapsto p_i^{k-n_i}@n_{min}\}$
    \STATE $\ca \gets \ca \uplus \{ax_c@n_{min}\}$ \COMMENT{add consecutive
      version of axiom}
    \STATE $\idxset \gets \idxset \uplus \{p_i^{k-n_i}@0, \dots, p_i^{k-n_i}@k\}$
    \STATE $\Abs{X} \gets X^p$; $\Abs{I} \gets I^p$; $\Abs{T} \gets T^p$; $\Abs{P} \gets P^p$
    \ENDFOR
    \STATE // Go through consecutive array axiom instantiations
    \FOR{$ax \in \ca$}
    \STATE \textbf{let} $n_{min} \coloneqq \mathit{min}(\tau(ax))$, $n_{max} \coloneqq \mathit{max}(\tau(ax))$
    \STATE assert($n_{max} = n_{min} \lor n_{max} = n_{min} + 1$)
    \IF{$k = 0$}
    \STATE $\Abs{I} \gets \Abs{I} \wedge ax\{X@n_{min} \mapsto X\}$ 
    \ELSIF{$n_{min} = n_{max}$} 
    \STATE $\Abs{T} \gets \Abs{T} \wedge ax\{X@n_{min} \mapsto X\} \wedge ax\{X@n_{min}
      \mapsto X'\}$ 
    \ELSE
    \STATE $\Abs{T} \gets \Abs{T} \wedge ax\{X@n_{min} \mapsto X\}\{X@(n_{min}+1)
    \mapsto X'\}$ 
    \ENDIF
    \ENDFOR
    \ENDLOOP
    \end{algorithmic}
\end{algorithm}

Line~1 computes an \define{index set} $\idxset$ of index terms with
$\ComputeIndices$ --- this set is used in the lazy axiom instantiation step
below. The procedure adds to $\idxset$ every term that appears as an index in a
$\Abs{\Readnoarg}$ or $\Abs{\Writenoarg}$ operation \iflongversion (recall that
these appear as uninterpreted functions in the abstracted STS and property) \fi
in $BMC(\Abs{\mathcal{S}}$, $\Abs{P}, k)$. Furthermore, it adds a witness index
for every array equality --- the witness corresponds to a Skolemized existential
variable in the contrapositive of axiom~\eqref{eq:array-ext}. For soundness, it
must also add an extra variable $\lambda_{\sigma}$ for each index sort $\sigma$
and constrain it to be different from all the other index variables of the same
sort (this is based on the approach in~\cite{bradley-arrays}). Intuitively, this
variable represents an arbitrary index different from those mentioned in the
STS.\@ We assume that the index sorts are from an infinite domain so that a
distinct element is guaranteed. For simplicity of presentation, we also assume
from now on that there is only a single index sort (e.g., integers). Otherwise,
$\idxset$ must be partitioned by sort. For the abstract STS in
Fig.~\ref{fig:abs-running}, with $k=1$, the index set would be $\idxset
\coloneqq \{\ridx@0, \widx@0, w_0@0, w_1@0, \lambda_{\Int}@0, \ridx@1, \widx@1,
w_0@1, w_1@1, \lambda_{\Int}@1\}$, where $w_0$ and $w_1$ are witness indices.

After computing indices, the algorithm enters the main loop. Line 3 dispatches a
bounded model checking query, $BMC(\Abs{\mathcal{S}}$, $\Abs{P}, k)$. The result
$\rho$ is either a counterexample, or the distinguished value $\bot$, indicating
that the query is unsatisfiable. If it is the latter, then it returns the
refined STS and property, as the property now holds on the STS up to bound $k$.
Otherwise, the algorithm continues. The next step (line~5) is to find violations
of array axioms in the execution $\rho$ based on the index set $\idxset$.

\ArrayAxiomsnoarg takes two arguments, a counterexample and an index set, and
returns instantiated array axioms that do not hold over the counterexample. This
works as follows. It first looks for occurrences of $\Abs{\Writenoarg}$ in the
BMC formula. For each such occurrence, it instantiates the
\eqref{eq:array-write} axiom so that the $\Abs{\Writenoarg}$ term in the axiom
matches the term in the formula (i.e., we use the $\Abs{\Writenoarg}$ term as a
trigger). This instantiates all quantified variables except for $i$. Next it
instantiates $i$ once for each variable in the index set. Each of the
instantiated axioms are evaluated using the values from the counterexample and
all instantiations that reduce to false are saved. The procedure does the same thing for the
\eqref{eq:array-const} axiom, using each constant array term in the BMC formula
as a trigger. Finally, for each array equality $a@m = b@n$ in the BMC formula,
it checks an instantiation of the contrapositive of \eqref{eq:array-ext}: $a@m
\not=b@n \to \Read{a@m}{w_i@n} \not= \Read{b@n}{w_i@n}$. The saved instantiated
formulas that do not hold in $\rho$ are added to the set of violated axioms.

\ArrayAxiomsnoarg sorts the collected, violated axiom instantiations into two
sets based on which timed variables they contain. The \emph{consecutive} set
contains formulas with timed variables whose timing differs by at most one;
whereas the timed variables in the formulas contained in the
\emph{non-consecutive} set may differ by more. Formally, let $\tau$ be a
function which takes a single timed variable and returns its time (e.g.,
$\tau(i@2) = 2$). We lift this to formulas by having $\tau(\phi)$ return the set
of all time-steps for variables in $\phi$. A formula $\phi$ is
\textit{consecutive} iff $\mathit{max}(\tau(\phi))-\mathit{min}(\tau(\phi)) \leq
1$. Note that instantiations of \eqref{eq:array-ext} are consecutive by
construction. Additionally, because constant arrays have the same value in all
time steps, the algorithm can always choose a representative time-step for instantiations
of \eqref{eq:array-const} that results in a consecutive formula. However,
instantiations of \eqref{eq:array-write} may be non-consecutive, because the
variable from the index set may be from a time-step that is different from that
of the trigger term. \ArrayAxiomsnoarg returns the pair $\langle \ca, \nca
\rangle$, where $\ca$ is a set of consecutive axiom instantiations and $\nca$ is
a set of pairs --- each of which contains a non-consecutive axiom instantiation
and the index-set term that was used to create that instantiation. We assume
that the index-set term used in a non-consecutive axiom is \emph{not} an
auxiliary variable. Since auxiliary variables only record or predict the value
of another index, it does not make sense to target one of these for prophecy.

Line 6 checks if the returned sets are empty. If so, then there are no
array axiom violations and $\rho$ is a concrete counterexample. In
this case, the system, property, and $\mathit{false}$ are returned. Otherwise,
the non-consecutive formulas are processed in lines 8--15.
Given a non-consecutive formula $ax$ together with its index-set variable
$i@n_i$, line 9 computes the minimum time-step of the axiom's other variables,
$n_{min}$. Then line 10 calls the \Prophecizenoarg method to create a prophecy variable
$p_i^{k-n_i}$, that is effectively a way to refer to $i@n_i$ at time-step
$n_{min}$ (line 10). This allows the algorithm to create a consecutive formula $ax_c$ that
is semantically equivalent to $ax$ (line 11). This new consecutive formula is
added to $\ca$ in line 12, and in line 13 the introduced prophecy variables (one
for each time-step) are added to the index set. Then, line 14 updates the
abstraction.

At line 17, the algorithm is left with a set of consecutive formulas to process. For each
consecutive formula $ax$, line 18 computes the minimum and maximum time-step of its
variables, which must differ by no more than 1 (line 19). There are
three cases to consider: i) when $k=0$, the counterexample consists of only the
initial state --- then the initial state is refined by adding the untimed version of
$ax$ to $\Abs{I}$ (line 21); ii) if $ax$ contains only variables from a single
time step, then the untimed version of $ax$ is added as a constraint for both $X$
and $X'$, ensuring that it will hold in every state (line 23); iii) finally, if
$ax$ contains variables from two adjacent time steps, it can be translated
directly into a transition formula to be added to $\Abs{T}$ (line 25). The loop
then repeats with the newly refined STS.

\paragraph{Example.} Consider again the example from Fig.~\ref{fig:abs-running},
and suppose $\RefineTSArrays$ is called on $\Abs{\mathcal{S}}$ and $\Abs{P}$
with $k=3$. At this unrolling, one possible abstract counterexample violates the
following nonconsecutive axiom instantiation: {\small
  \begin{equation*}
    \begin{aligned}
      (\ridx@2 = \widx@0 & \implies
      \AbsRead{\AbsWrite{a@0}{\widx@0}{\data_w@0}}{\ridx@2} = \data_w@0) ~ \wedge \\
      (\ridx@2 \neq \widx@0 & \implies
      \AbsRead{\AbsWrite{a@0}{\widx@0}{\data_w@0}}{\ridx@2} = \AbsRead{\Abs{a@0}}{\ridx@2})
    \end{aligned}
  \end{equation*}
}

\noindent
To make this nonconsecutive axiom consecutive, it introduces a prophecy variable.
The target will be the instantiated index, $\ridx@2$. The relevant value to
predict is one step before a possible property violation (i.e., the end of a
finite path), because $k=3$, and $\tau(\ridx@2) = 2$, thus $k-\tau(\ridx@2) =
1$. This corresponds to the $k - n_i$ at line 10 of Algorithm~\ref{alg:refine-ts}.
Calling $\Prophecize{\Abs{\mathcal{S}}}{\Abs{P}}{\ridx}{1}$
returns the new STS $\langle \langle \Abs{X} \uplus \{\hist{\ridx}{1},
p_{\ridx}^1\}, \Abs{I}, \Abs{T} \wedge \hist{\ridx}{1'} = \ridx \wedge
p_{\ridx}^{1'} = p_{\ridx}^{1} \rangle$ and the new property $p_{\ridx}^1 =
\hist{\ridx}{1} \implies \data_r < 200$. The history variable $\hist{\ridx}{1}$
makes the previous value of $\ridx$ available at each time-step, and the
prophecy variable $p_{\ridx}^1$ predicts the value of $\ridx$ one step before a
possible property violation. The axiom will be updated by replacing $\ridx@2$
with the prophecy variable which has the same value. Since the prophecy variable
is frozen, it is the same at every step. Thus, it can choose the prophecy
variable at a time-step that makes the axiom consecutive. In this case, the algorithm
substitutes $p_{\ridx}^1@0$ for $\ridx@2$. This results in the following
consecutive axiom: {\small
  \begin{equation*}
    \begin{aligned}
      (p_{\ridx}^1@0 = \widx@0 & \implies
      \AbsRead{\AbsWrite{a@0}{\widx@0}{\data_w@0}}{p_{\ridx}^1@0} = \data_w@0) ~ \wedge \\
      (p_{\ridx}^1@0 \neq \widx@0 & \implies
      \AbsRead{\AbsWrite{a@0}{\widx@0}{\data_w@0}}{p_{\ridx}^1@0} = \AbsRead{\Abs{a@0}}{p_{\ridx}^1@0})
    \end{aligned}
  \end{equation*}
}

The untimed version (and a primed version) of this consecutive axiom would be
added to the transition relation at line 23 of Algorithm~\ref{alg:refine-ts}.

\smallbreak We stress that processing nonconsecutive axioms using
$\Prophecizenoarg$ is how Algorithm~\ref{alg:refine-ts} automatically discovers
the universal prophecy variable $p_{\ridx}^{1}$, and it is exactly the universal
prophecy variable that was needed in Section~\ref{sec:motivation} to prove
correctness of the running example. An alternative approach could avoid
nonconsecutive axioms using Craig interpolants~\cite{craig-interpolation} so
that only consecutive axioms are found~\cite{euficient}. However,
quantifier-free interpolants are not guaranteed to exist for the standard theory
of arrays~\cite{itp-dt,diff-array-interp}, and the auxiliary variables found
using nonconsecutive axioms are needed to improve the chances of finding a
quantifier-free inductive invariant. It is thus extremely important to start
with a weak abstraction that allows us to examine spurious counterexamples in
the BMC unrolling and find nonconsecutive axiom instantiations, which are then
used to identify good prophecy targets.

\subsection{Correctness}
We now state two important correctness theorems.
\iflongversion
\else
  Note that here and below, proofs are omitted due to space constraints. An extended version with
  proofs is available at: \url{https://arxiv.org/abs/2101.06825}.
\fi

\begin{thm}
  \label{thm:soundness}
  Algorithm~\ref{alg:proph-array}, instantiated with $\AbstractArrays$, a
  sound model-check\-er $\Provenoarg$ as described above, and $\RefineTSArrays$ is
  sound, i.e., if it returns \emph{true} then the property does hold.
\end{thm}

\iflongversion
  \noindent\textit{Proof Sketch.}
  Algorithm~\ref{alg:proph-array} only returns true if \Provenoarg succeeds in
  proving the property. Our initial abstraction only removes the array theory
  semantics, but leaves every other theory intact, so it is a sound abstraction.
  The refinement performed by $\RefineTSArrays$ is also sound.
  \Prophecizenoarg first optionally applies \Delaynoarg depending on the input
  arguments, then introduces a prophecy variable. Theorem~\ref{thm:hist}
  guarantees that \Delaynoarg preserves the invariance of the property.
  Furthermore, introducing a prophecy variable is accomplished by directly
  applying Theorem~\ref{thm:intro-univ}, followed by Theorem~\ref{thm:prophecy},
  which additionally guarantee that the resulting system and property are
  invariant if and only if the original property is invariant on the original
  system. Thus, the entire \Prophecizenoarg procedure produces a new system and
  property that preserve invariance with respect to the initial query. Finally,
  each axiom instantiation is, by definition, valid in the theory of arrays, and
  lifting them simply requires them to hold in every state of a path.
  Furthermore, this does not rule out any true counterexamples, as the
  interpretations in true counterexamples must be $\T_A$-interpretations.
  Therefore, if at any point $\Provenoarg$ is able to prove the property, it
  follows that the original property holds on the original concrete system,
  $\mathcal{S}$. \hfill$\Box$ \fi

\begin{thm}
  \label{thm:concrete-counterexamples}
  If Algorithm~\ref{alg:proph-array}, instantiated with $\AbstractArrays$,
  $\Provenoarg$ as described above, and $\RefineTSArrays$, returns false, there
  is a counterexample in the concrete transition system.
\end{thm}

\iflongversion
  \noindent\textit{Proof Sketch.}
  Theorems~\ref{thm:hist},~\ref{thm:prophecy}, and~\ref{thm:intro-univ} ensure
  that invariance of the property is preserved when adding auxiliary variables.
  Algorithm~\ref{alg:proph-array} returns false only when \RefineTSArrays
  returns false in line 6 of Algorithm~\ref{alg:refine-ts}. This occurs if the
  refinement procedure is unable to find any array axioms that are violated in
  the BMC formula ($\rho$ from line 3 of Algorithm~\ref{alg:refine-ts}). It
  suffices to prove that if all enumerated axioms hold, then the BMC formula is
  satisfiable in $\T_A$ and there is a length $k$ counterexample.

  The \define{array property
    fragment}~\cite{bradley-arrays,calculuscomp,decision-procedures} is a
  fragment of the theory of arrays that allows some universal quantification
  while staying decidable. It is defined for an extensional theory of arrays
  with $\Readnoarg$ and $\Writenoarg$ functions with the semantics given by
  \eqref{eq:array-write} and \eqref{eq:array-ext}. The fragment is of the form
  $\forall \vec{i}~.~ \phi_I(\vec{i}) \rightarrow \phi_V(\vec{i})$, for a vector
  of bound index variables $\vec{i}$, index guard $\phi_I$, and value constraint
  $\phi_V$. Both $\phi_I$ and $\phi_V$ are constrained by a grammar. Our BMC
  queries are quantifier-free which falls within the array property fragment.
  The only universal quantifiers are hidden by the \eqref{eq:array-const} axiom
  (because constant arrays are not explicitly included in the array property
  fragment theory of arrays). However, this simple form of universal
  quantification is contained in the fragment. Thus, our queries are a strict
  subset of the more general array property fragment.

  Our axiom enumeration is based on a reduction
  technique~\cite{bradley-arrays,calculuscomp,decision-procedures} for the array
  property fragment that is sound and complete. Because the technique is
  complete, if the abstract formula is satisfiable and all enumerated axioms are
  true, then the original $\T_A$ formula is satisfiable. \hfill$\Box$

\fi

\section{Expressiveness and Limitations}
\label{sec:expressiveness}

\subsection{Expressiveness}
We now address the expressiveness of counterexample-guided prophecy with regard
to the introduction of auxiliary variables. For simplicity, we ignore the array
abstraction, relying on the correctness theorems. An inductive invariant using
auxiliary variables can be converted to one without auxiliary variables by first
universally quantifying over the prophecy variables, then existentially
quantifying over the history variables. The details are captured by this
theorem:

\begin{thm}
  \label{thm:expressiveness}
  Let $\mathcal{S} \coloneqq \langle X, I, T \rangle$ be an STS, and $P(X)$ be a
  property such that $\mathcal{S} \models P(X)$. Let $H$ be the set of history
  variables, and $\mathcal{P}$ be the set of prophecy variables introduced by
  $\RefineTSArrays$. Let $\tilde{\mathcal{S}} \coloneqq \langle X \cup H \cup
  \mathcal{P}, I, \tilde{T} \rangle$ and $\tilde{P} \coloneqq (\bigwedge_{p \in
    \mathcal{P}} p = \tilde{t}(p)) \implies P(X)$ be the system and property
  with auxiliary variables. The function $\tilde{t}$ maps prophecy variables to
  their target term from \Prophecizenoarg. If $\Inv(X, H, \mathcal{P})$ is an
  inductive invariant for $\tilde{\mathcal{S}}$ and entails $\tilde{P}$, then
  $\exists H \forall \mathcal{P} Inv(X, H, \mathcal{P})$ is an inductive
  invariant for $\mathcal{S}$ and entails $P$, where $\exists H$ and $\forall
  \mathcal{P}$ bind each variable in the set with the corresponding quantifier.
\end{thm}

\iflongversion
\begin{proof}
  We assume that $\Inv(X, H, \mathcal{P})$ is an inductive invariant that
  guarantees $\tilde{P}$. Equivalently, it meets the following conditions:

  \begin{equation}
    \tag{initiation}
    \label{eq:aux-init}
    \tilde{I} \models \Inv(X, H, \mathcal{P})
  \end{equation}
  \begin{equation}
    \tag{consecution}
    \label{eq:aux-ind}
    \Inv(X, H, \mathcal{P}) \wedge \tilde{T} \models \Inv(X', H', \mathcal{P}')
  \end{equation}
  \begin{equation}
    \tag{safety}
    \label{eq:aux-safety}
    \Inv(X, H, \mathcal{P}) \models \tilde{P}
  \end{equation}
  We must show that $\exists H \forall \mathcal{P} \Inv(X, H, \mathcal{P})$ is
  an inductive invariant of $\mathcal{S}$ and entails $P$. We accomplish this by
  demonstrating that each of the three conditions must hold.

  \textbf{Initiation:} $I \models \exists H \forall \mathcal{P} \Inv(X, H,
  \mathcal{P})$. This holds trivially because $I$ is unchanged, i.e., no
  auxiliary variables appear in the initial state constraint.

  \textbf{Consecution:} $\exists H \forall \mathcal{P} \Inv(X, H, \mathcal{P})
  \wedge T \models \exists H' \forall \mathcal{P}' \Inv(X', H', \mathcal{P}')$.
  This is equivalent to the following formula (manipulated into negation normal
  form) being unsatisfiable:
  \begin{equation}
    \label{eq:quant-consec}
    \exists H \forall \mathcal{P} \Inv(X, H, \mathcal{P}) \wedge T \wedge
    \forall H' \exists \mathcal{P}' \neg \Inv(X', H', \mathcal{P}')
    \end{equation}

     To complete this
    part of the proof, we introduce the function $\tilde{\sigma}$ which maps
    primed history variables to their next state update term from \Delaynoarg.
    For example, suppose we called $\Delay{\mathcal{S}}{x}{2}$ for state
    variable $x$, then $\tilde{\sigma}(h_x^{1'}) = x$ and
    $\tilde{\sigma}(h_x^{2'}) = h_x^1$. Crucially, the terms in the range of
    $\tilde{\sigma}$ do not contain variables from $P$, because prophecy
    variables are not targeted by $\Prophecizenoarg$.
    With this notation, the consecution of $\Inv(X, H, \mathcal{P})$ for
    $\tilde{T}$ means that the following formula is unsatisfiable:
    \begin{equation}
      \label{eq:consec-unsat}
    \Inv(X, H, \mathcal{P}) \wedge T \wedge \left( \bigwedge_{h \in H} h' = \tilde{\sigma}(h') \right) \wedge \left( \bigwedge_{p \in \mathcal{P}} p' = p \right) \wedge
    \neg \Inv(X', H', \mathcal{P}')
      \end{equation}
    We now show that the fact that \eqref{eq:consec-unsat} is unsatisfiable
    entails that \eqref{eq:quant-consec} is unsatisfiable.
    First, observe that \eqref{eq:consec-unsat} is equisatisfiable with
    \begin{equation}
      \label{eq:consec-subst}
    \Inv(X, H, \mathcal{P}') \wedge T \wedge \neg \Inv(X', \tilde{\sigma}(H'), \mathcal{P}')
      \end{equation}
      Next, if \eqref{eq:quant-consec} is satisfiable, then the following
      formula is satisfiable, where we drop the quantifiers over $H$ and replace
      them with fresh uninterpreted constants (for convenience, we simply drop
      the quantifiers and treat the free variables as uninterpreted constants):
      $\forall \mathcal{P} \Inv(X, H, \mathcal{P}) \wedge T \wedge \forall H'
      \exists \mathcal{P}' \neg \Inv(X', H', \mathcal{P}')$. If this formula is
      satisfiable, then the following formula is also satisfiable, where we
      instantiate the universal quantifier over $H'$ with $\tilde{\sigma}(H')$:
      $\forall \mathcal{P} \Inv(X, H, \mathcal{P}) \wedge T \wedge 
      \exists \mathcal{P}' \neg \Inv(X', \tilde{\sigma}(H'), \mathcal{P}')$.
      Finally, if this formula is satisfiable, then we can drop the existential quantifiers
      for $P'$ and instantiate the universal quantifier for $P$ with $P'$, which
      gives that \eqref{eq:consec-subst} is satisfiable. Since
      \eqref{eq:consec-subst} is unsatisfiable, then \eqref{eq:quant-consec}
      must be unsatisfiable as well.

  \textbf{Safety:} $\exists H \forall \mathcal{P} \Inv(X, H, \mathcal{P})
  \models P(X)$. This holds when $\exists H \forall \mathcal{P} \Inv(X, H,
  \mathcal{P}) \wedge \neg P(X)$ is unsatisfiable. To show this, we construct
  quantifier instantiations such that the resulting formula must be
  unsatisfiable. We first instantiate all the variables in $H$ with fresh
  constants by dropping the existential quantification. Next, we instantiate each
  of the universally quantified $p \in \mathcal{P}$ with their target term
  $\tilde{t}(p)$. Note that this target term might be a history variable from
  $H$ which is now instantiated. We allow $\tilde{t}$ to be applied to sets in
  the straightforward way. A model for the resulting formula, $\Inv(X, H,
  \mathcal{P})\{\mathcal{P} \mapsto \tilde{t}(\mathcal{P})\} \wedge \neg P(X)$,
  would be a counterexample for assumption \eqref{eq:aux-safety}. Thus it must
  be unsatisfiable.

  We have shown that initiation holds trivially, and that consecution and safety
  hold using quantifier instantiations. Thus, $\exists H \forall \mathcal{P}
  \Inv(X, H, \mathcal{P})$ must be an inductive invariant for $\mathcal{S}$ and
  $P(X)$.
  \end{proof}
\fi

Although the invariants found using counterexample-guided prophecy correspond to
$\exists \forall$ invariants over the unmodified system, we must acknowledge
that the existential power is very weak. The existential quantifier is only used
to remove history variables. While history variables can certainly be employed for
existential power in an invariant~\cite{temporal-prophecy}, these specific
history variables are introduced solely to target a term for
prophecy and only save a term for some fixed, finite number of steps. Thus, we
do not expect to gain much existential power in finding invariants on practical
problems. This use of history and prophecy variables can be
thought of as quantifier instantiation at the model checking level, where the
instantiation semantically uses a term appearing in an execution of the system.
Consequently, our technique performs well on systems where there is only a small
number of instantiations needed over terms that are not too distant in time from
a potential property violation that must be disproved (i.e., not many history
variables are required). This appears to be a common situation for
invariant-finding benchmarks, as we show empirically in Section~\ref{sec:experiments}.

\subsection{Limitations}
If our CEGAR loop terminates, it either terminates with a
proof or with a true counterexample. However, it is possible that the procedure
may not terminate. In particular, while we can always refine the abstraction for
a given bound $k$, there is no guarantee that this will eventually result in a
refinement that rules out all spurious counterexamples (of any length).

This failure mode occurs, for instance, when no finite number of calls to
$\Prophecizenoarg$ can capture all the relevant indices of the array. Consider
an example system with $I \coloneqq a \arreq \constarr{0}$, $T \coloneqq a'
\arreq \Write{a}{i_0}{\Read{a}{i_1} + 1}$, and $ P \coloneqq \Read{a}{\ridx}
\geq 0$. The array $a$ is initialized with 0 at every index, and at every step,
$a$ is updated at a single index by reading from an arbitrary index of $a$ and
adding 1 to the result.\footnote{An even simpler system which does not add 1 in
  the update would already be problematic; however, for that case, it is
  straightforward to extend our algorithm to have it learn that the array does
  not change.} Note that the index variables are unconstrained: they
can range over the integers freely at each time step. The property is that
reading from $a$ at $i_r$ returns a positive value. This property holds because
of a quantified invariant maintained by the system: $\forall i~.~\Read{a}{i}
\geq 0$.

However, the initial abstraction is a memoryless array which can easily violate
the property by returning negative values from reads. Since the array is updated
in each step at an arbitrary index based on a read from another arbitrary index,
no finite number of prophecy variables (of the form used in $\Prophecizenoarg$)
can capture all the relevant indices. It will successively rule out longer
finite spurious counterexamples, but will never be refined enough to prove the
property unboundedly. Note that this is related to our abstraction and choice to
limit prophecy to predicting values a fixed, finite number of steps before a
potential property violation. Another form of prophecy variable could be used to
prove this property. For example, a prophecy variable that predicts the first
index value that stores a negative value in $a$ could be used to show that this
cannot happen.

We believe that this issue can be circumvented in an automated fashion with
future work. In fact, an approach introduced since the conference
version~\cite{cegp} of this paper uses prophecy variables with a different
refinement loop for verifying parameterized protocols, which cannot be handled
by our technique due to this limitation~\cite{proph-param}.

A related, but less fundamental issue is that the index set might not contain
the best choice of targets for prophecy. While the index set \emph{is} sufficient for
ruling out bounded counterexamples, it is possible there is a better target for
universal prophecy that does not appear in the index set. However, based on the
evaluation in Section~\ref{sec:experiments}, it appears that the index set does
work well in practice.

\section{Implementation Details}
\label{sec:prototype}

We will now describe our prototype of counterexample-guided prophecy along with
some practical implementation details. Recall that
Algorithm~\ref{alg:proph-array} can use any unbounded model checking technique for
$\Provenoarg$. In our prototype, we choose to instantiate it with
\icthreeia~\cite{ic3ia-code} (downloaded Apr 27, 2020), an open-source C++
implementation of IC3 via Implicit Predicate Abstraction (IC3IA)~\cite{ic3ia},
which is itself a CEGAR loop that uses implicit predicate abstraction to perform
IC3~\cite{ic3,ctigar} on infinite-state systems and uses interpolants to find
new predicates. \icthreeia uses MathSAT~\cite{mathsat5} (version 5.6.3) as the
backend SMT solver and interpolant producer. We call our prototype
\prophic~\cite{prophic3-prototype}.

\subsection{Engineering Heuristics and Options}

\subsubsection*{Weak and Strong Abstraction.}
It is important to have enough prophecy variables to assist in constructing
inductive invariants. We found that we could often obtain a larger, richer set
of prophecy variables by weakening our array abstraction. We do this by
replacing equality between arrays by an uninterpreted predicate, and also
checking the congruence axiom, the converse of \eqref{eq:array-ext}. Since more
axioms are checked, there are more opportunities to introduce auxiliary
variables. We call this \emph{weak} abstraction (\WA) as opposed to
\emph{strong} abstraction (\SA), which uses regular equality between abstract
arrays and guarantees congruence through UF axioms. Our default configuration
uses weak abstraction.

\subsubsection*{Lemma and Auxiliary Variable Filtering.}
Although the algorithm depends on introducing auxiliary variables, an excessive
number of unnecessary auxiliary variables could overwhelm the $\Provenoarg$
step. Thus, an improvement not shown in Algorithm~\ref{alg:refine-ts} is to
check consecutive axioms first and only add nonconsecutive ones when necessary.
This is the motivation behind the custom array solver implementation
$\ArrayAxiomsnoarg$ based on~\cite{bradley-arrays}. In principle, we could have
used an SMT solver to find array axioms, but it would give no preference to
consecutive axioms. Even when enumerating consecutive axioms first, we can still
end up with more auxiliary variables than necessary. We use an unsat-core based
procedure to prune nonconsecutive refinement axioms. In particular, we attempt
to remove nonconsecutive axioms that target indices at times further from the
end of the trace, because they would introduce more history variables. In
practice, this can substantially reduce the number of added auxiliary variables.

Similarly, we could overwhelm the algorithm with unnecessary consecutive axioms.
$\ArrayAxiomsnoarg$ can still produce hundreds or even thousands of
(consecutive) axiom instantiations. Once these are lifted to the transition
system, some may be redundant. To mitigate this issue, when the BMC check
returns $\bot$ and we are about to return (line 4 of
Algorithm~\ref{alg:refine-ts}), we keep only axioms that appear in the unsat core of the
BMC formula~\cite{DBLP:journals/jair/CimattiGS11}.

\subsubsection*{Abstract Values Refinement Loop.}
In our implementation, we also include a simple abstraction-refinement wrapper
which abstracts large constant integers and refines them with the actual values
if that fails. This is especially useful for verifying software benchmarks with
large constant loop bounds. Otherwise, the system might need to be unrolled to a
very large bound to reach an abstract counterexample. This was only necessary
for a handful of benchmarks in the first benchmark set.

\subsubsection*{Assume Property in Pre-State.}
As long as we are only interested in the \textit{first} violation of a property,
we can assume that the property was not violated in the past. This observation
is formalized in Theorem~5 of~\cite{eager-abstraction}. It is common to achieve
this for invariant checking by assuming the property over current state
variables in the transition relation, so that every transition starts in a state
satisfying the property.

In the context of counterexample-guided prophecy, this strategy may prove useful
because the property is weakened with each call to
Algorithm~\ref{alg:prophecize} (the original property becomes the consequent of an
implication). We can assume the original (stronger) property in all previous
states which can help the algorithm converge.

\begin{figure}[t]
  \centering
\begin{lstlisting}[
  boxpos=t,
  language=C++,
  stepnumber=1,
  numberstyle=\color{gray},
  basicstyle=\ttfamily\small,
  keywords={return,false,true,for},
  keywordstyle=\color{blue}\ttfamily,
  stringstyle=\color{mauve}\ttfamily,
  commentstyle=\color{dkgreen}\ttfamily,
  classoffset=2,
  morekeywords={int,assert,assume},
  keywordstyle=\color{burgundy}\ttfamily\bfseries,
  ]

  int N;
  assume ( N > 0 );

  int a[N];
  int c;
  for ( int i = 0 ; i < N ; i++ ) {
    a[i] = c;
  }

  for ( int j = 0 ; j < N ; j++ ) {
    assert( a[j] == c );
  }
  \end{lstlisting}
  \caption{Initialize Array. \label{fig:c-array-init-const}}
\end{figure}

\begin{figure}[t]
  \small
  \begin{equation*}
    \begin{aligned}
      I \coloneqq \ & i = 0 \wedge j = 0 \wedge \neg \err \\
      T \coloneqq \ & i < N \rightarrow i' = i + 1 \ \wedge \\
                  & i < N \rightarrow a' = \Write{a}{i}{c} \ \wedge \\
                  & i < N \rightarrow j' = j \ \wedge \\
                  & i < N \rightarrow \err' = \err \ \wedge \\
                  & (i \geq N \wedge j < N) \rightarrow i' = i \ \wedge \\
                  & (i \geq N \wedge j < N) \rightarrow a' = a \ \wedge \\
                  & (i \geq N \wedge j < N) \rightarrow j' = j + 1 \ \wedge \\
                  & (i \geq N \wedge j < N \wedge \Read{a}{j} = c \wedge \neg \err) \rightarrow \neg \err' \ \wedge \\
                  & (i \geq N \wedge j \geq N) \rightarrow \bot \\
      P \coloneqq \ & \neg \err
    \end{aligned}
  \end{equation*}
  \caption{Possible STS encoding of Fig.~\ref{fig:c-array-init-const}.
    \label{fig:sts-array-init-const}}
  \vspace{-1em}
\end{figure}

Consider the C program shown in Fig.~\ref{fig:c-array-init-const}, based on
one of the benchmarks evaluated in Section~\ref{sec:experiments}. The example
populates an array of arbitrary size $N$, with an arbitrary, fixed, value $c$.
Fig.~\ref{fig:sts-array-init-const} shows one possible encoding of this
program as an STS.\@ This encoding is carefully chosen to illustrate a case where
assuming the property in the pre-state is needed for counterexample-guided
prophecy to converge. Other possible encodings could avoid this issue entirely.

In this encoding, if we do not assume the original property in the pre-state, we
observed that \prophic would diverge and introduce an increasing number of
prophecy variables. Consider a case where $N$ is assigned a specific value, $5$.
Since the array abstraction starts memoryless, the algorithm needs to add 5
prophecy variables to refine the memory. However, since $N$ is arbitrary, this
results in an infinite chain of new prophecy variables as longer traces are
considered. Furthermore, each time a prophecy variable is introduced, the
underlying IC3IA algorithm is restarted with a weaker property. This means that
the original property, $\neg \err$, cannot be assumed in the pre-state. Note,
the new property produced by $\Prophecizenoarg$ is an implication that will be
trivially true in most of the path. The most important part of the transition
relation in Fig.~\ref{fig:sts-array-init-const} to consider is the
second-to-last line of $T$. Let that be the \emph{error rule}. Since we do not
assume $\neg \err$, the error rule might be trivially satisfied. Intuitively,
without this assumption, we need to justify the assertion for the value of index
$j$ at every time-step.

However, we know it is safe to assume the original property in the pre-state for
counter-example guided prophecy. If we do so, the algorithm converges. This is
because that assumption coupled with predicting a single $j$, one step before a
potential property violation, is sufficient to enforce the error rule which
ensures $\neg \err$ holds in the next state.

\subsubsection*{Important Variables.}
In many implementations of IC3IA, including \icthreeia, new predicates are
obtained by mining interpolants from an unsatisfiable spurious counterexample
trace conjoined with the concrete unrolled transitions. Typically not all of
these predicates are necessary, so they are often reduced using unsatisfiable
cores. However, in the context of counterexample-guided prophecy, we might
prefer certain predicates. In particular, predicates involving prophecy
variables are good candidates, since we know the prophecy variable was necessary
to rule out a spurious counterexample. Note that there are two levels of
abstraction refinement in this context: the array abstraction and refinement for
counterexample-guided prophecy, and the predicate abstraction in IC3IA.\@ Here, we
are focused on the latter. One heuristic we tried is always keeping predicates
that use a prophecy variable.

\subsubsection*{Finite-domain Indices.}
In this paper, we have assumed that the index sort has an infinite domain. This
fits our domain of problems that require quantified invariants. If the sort is
finite, a universal quantifier could in principle be enumerated by instantiating
it with every possible value. Although, it is certainly possible that a
quantifier is much more efficient than this enumeration.

The restriction to infinite domain indices is also a technical limitation of the
array solving technique of~\cite{bradley-arrays}, which our approach is based
on. This restriction is shared by many SMT solvers, particularly when there
are chains of equalities between writes on constant arrays with different bases,
e.g., $
      a = \constarr{0_8} \wedge
      b = \constarr{1_8} \wedge
      \Write{a}{i_2}{e_8} = \Write{b}{j_2}{d_8}$, where $c_w$ is a bitvector variable or value $c$ with width $w$.
An infinite domain allows us to assume that there is always an index value that
has not been used in the array formula. This is crucial for the $\lambda$ index,
which has the primary goal of referring to indices initialized by the constant
array axiom that were never overwritten. In a finite domain, we cannot make this
assumption. See~\cite{bradley-arrays,calculuscomp} for more information on the
$\lambda$ index.

This limitation is only with regards to the array axiom enumeration. The other
contributions of this paper, including using prophecy variables in place of
universal quantification, are entirely applicable over finite domains. One
low-effort approach for applying counterexample-guided prophecy over
finite-domain indices is to give up completeness. By simply not including axioms
over a $\lambda$ index for finite-domain sorts, the array solving procedure
might conclude the query is satisfiable when it is actually unsatisfiable. Thus,
the overall algorithm could return spurious counterexamples, but would still
soundly return proofs. A given spurious counterexample is finite and would be
straightforward to analyze, either with a dedicated checker or an SMT solver
without this limitation.

\section{Experiments}
\label{sec:experiments}

\subsection{Setup}
We evaluate our tool against three state-of-the-art tools
for inferring universally quantified invariants over linear arithmetic and
arrays: \freqhorn, \quicthree, and \gspacer. All these tools are Constrained
Horn Clause (CHC) solvers built on Z3~\cite{z3}. The algorithm implemented in
this version of
\freqhorn~\cite{freqhorn-implementation} is a \emph{syntax-guided
synthesis}~\cite{sygus} approach for inferring universally quantified
invariants over arrays~\cite{freqhorn}. \quicthree is built on
Spacer~\cite{spacer}, the default CHC engine in Z3, and extends IC3 over linear
arithmetic and arrays to allow universally quantified frames (frames are
candidates for inductive invariants maintained by the IC3 algorithm)~\cite{quic3}. It also
maintains a set of quantifier instantiations which are provided to the
underlying SMT solver. \texttt{quic3} was recently incorporated into Z3. We used
Z3 version 4.8.9 with parameters suggested by the \texttt{quic3}
authors.\footnote{\texttt{fp.spacer.q3.use\_qgen=true} \texttt{fp.spacer.ground\_pobs=false}\\
  \texttt{fp.spacer.mbqi=false} \texttt{fp.spacer.use\_euf\_gen=true}} Finally,
\gspacer is an extension of \texttt{Spacer} which adds three new inference rules for
improving local generalizations with global guidance~\cite{gspacer}. While this last technique
does not specifically target universally quantified invariants, it can be used
along with the \quicthree options in \texttt{Spacer} and potentially executes a much
different search. The \gspacer submission~\cite{gspacer-submission} won the
arrays category in CHC-COMP 2020~\cite{chc-comp-2020-report}. We use the same
configuration entered in the competition. We also include
\icthreeia and the default configuration of \texttt{Spacer} in our results, neither of
which can produce universally quantified invariants. Our default configuration
of \prophic uses weak abstraction. We chose to build our prototype on
\icthreeia instead of \texttt{Spacer}, in part because we needed uninterpreted functions
for our array abstraction, and \texttt{Spacer} does not handle them in a straightforward
way, due to the semantics of CHC~\cite{horn-solvers}.

We compare these solvers on four benchmark sets: i) \textit{freqhorn} --- all
benchmarks containing arrays in~\cite{freqhorn-benchmarks} from
the \freqhorn paper~\cite{freqhorn}; ii) \textit{quic3} --- benchmarks from the
\quicthree paper~\cite{quic3} (these were C programs from SV-COMP~\cite{sv-comp}
that were modified to require universally quantified invariants); iii)
\textit{vizel} --- additional benchmarks provided to us by the authors of~\cite{quic3};
and iv) \textit{chc-comp-2020} --- the array category benchmarks of
CHC-COMP 2020~\cite{chc-comp-2020} (as explained below, these contain a
translation of the \textit{quic3} benchmarks). Additionally, we sort the
benchmarks into four categories: 1) Q --- safe benchmarks solved by some tool
supporting quantified invariants but none of the solvers that do not; 2) QF ---
those solved by at least one of the tools that do not support quantified
invariants, plus any unsafe benchmarks; 3) US --- unsafe benchmarks and 4) UK ---
unknown (i.e., unsolved) benchmarks. Because not all of the benchmark sets were
guaranteed to require quantifiers, this is an approximation of which benchmarks
required quantified reasoning to prove safe.

Both \prophic and \icthreeia take a transition system and property specified in
the Verification Modulo Theories (VMT) format~\cite{vmt-lib}, which is a
transition system format built on SMT-LIB~\cite{smt-lib}. All other solvers read
the CHC format. We translated benchmark sets \textit{freqhorn} and \textit{chc-comp-2020} from CHC to VMT using the
\emph{horn2vmt} program which is distributed with \icthreeia. For benchmark sets
\textit{quic3} and \textit{vizel}, we started with the C programs and generated both VMT and CHC using
\emph{Kratos2} (an updated version of \emph{Kratos}~\cite{kratos}). We note that
\textit{chc-comp-2020} includes another translation of the \textit{quic3} benchmarks to CHC
by \texttt{SeaHorn}~\cite{seahorn}.
We ran all experiments on a 3.5GHz Intel Xeon E5-2637 v4 CPU with a timeout of 2
hours and a memory limit of 32GB.
\iflongversion
\else
An artifact for reproducing these results is publicly available~\cite{artifact,tacas21-vm}.
\fi

\iflongversion
\begin{figure}
  \includegraphics[width=\linewidth]{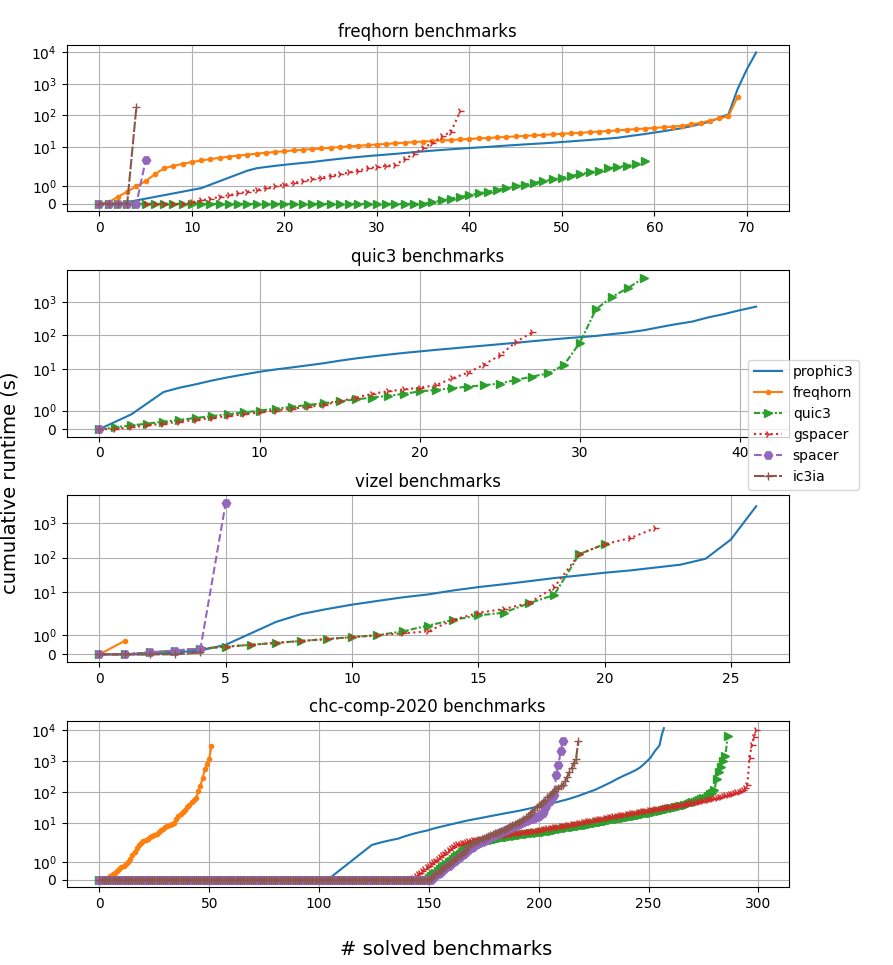}
  \caption{Number of solved benchmarks over time (sorted).}
  \label{fig:cactus}
\end{figure}
\fi

\begin{figure}[t]
  \centering
  \setlength\tabcolsep{0.2em}
  \begin{tabularx}{\linewidth}{X c c c c c}
    \toprule
    \bfseries solver & freqhorn (81) & quic3 (43) &
    vizel (33) & chc-comp-2020 (501) & \bfseries total \\
    \midrule
    \prophic & 66/5/0 & 41/0/0 & 22/3/1 & 42/159/56 & 171/167/57 \\
    \freqhorn & 65/4/0 & 0/0/0 & 0/1/0 & 4/46/1 & 69/51/1 \\
    \quicthree & 55/4/0 & 34/0/0 & 15/4/1 & 74/137/75 & 178/145/76 \\
    \gspacer & 34/5/0 & 27/0/0 & 18/3/1 & 66/139/94 & 145/147/95 \\
    \texttt{Spacer} & 0/5/0 & 0/0/0 & 0/4/1 & 0/134/77 & 0/143/78 \\
    \icthreeia & 0/4/0 & 0/0/0 & 0/3/1 & 0/158/60 & 0/165/61 \\
    \bottomrule
  \end{tabularx}
  \caption{Experimental results. They are reported as \emph{\#~Q} /
    \emph{\#~QF} / \emph{\#~US}. \label{fig:reorganized-results}}
\end{figure}

\begin{figure}[t]
  \centering
  \setlength\tabcolsep{0.3em}
  \begin{tabularx}{\linewidth}{X c c c c c}
    \toprule
    \bfseries solver & freqhorn (81) & quic3 (43) & vizel (33) & chccomp2020 (501) &
    \bfseries total \\
    \midrule
    \prophic & 66/5/0 & 41/0/0 & 22/3/1 & 42/159/56 & 171/167/57 \\
    \texttt{prophic3\_sa} & 62/5/0 & 38/0/0 & 20/3/1 & 41/159/65 & 161/167/66 \\
    \texttt{prophic3\_nav} & 57/4/0 & 42/0/0 & 21/3/1 & 43/160/55 & 163/167/56 \\
    \texttt{prophic3\_na} & 68/4/0 & 41/0/0 & 18/3/1 & 44/159/56 & 171/166/57 \\
    \texttt{prophic3\_npr} & 66/4/0 & 42/0/0 & 20/3/1 & 45/159/56 & 173/166/57 \\
    \texttt{prophic3\_ntp} & 66/4/0 & 40/0/0 & 20/3/1 & 33/159/56 & 159/166/57 \\
    \texttt{prophic3\_nur} & 66/5/0 & 42/0/0 & 20/3/1 & 40/159/55 & 168/167/56 \\
    \texttt{prophic3\_nhp} & 68/4/0 & 41/0/0 & 22/3/1 & 42/159/56 & 173/166/57 \\
    \texttt{prophic3\_nar} & 64/4/0 & 42/0/0 & 21/3/1 & 42/159/55 & 169/166/56 \\
    \texttt{prophic3\_noheur} & 65/5/0 & 38/0/0 & 13/3/1 & 28/158/57 & 144/166/58 \\
    \bottomrule
  \end{tabularx}
  \caption{Self-comparison with different options, reported as \emph{\#~Q} /
    \emph{\#~QF} / \emph{\#~US}. \label{fig:self-comparison}}
\end{figure}

\subsection{Results}
The results are shown in Fig.~\ref{fig:reorganized-results} as a table.
\iflongversion Fig.~\ref{fig:cactus} shows cactus plots demonstrating the number of solved
benchmarks over time.\fi
We first observe that \prophic solves the most benchmarks in
the \textit{freqhorn}, \textit{quic3}, and \textit{vizel} benchmark sets, both overall and in category Q. The \textit{quic3} (and
most of the \textit{freqhorn}) benchmarks require quantified invariants; thus,
\icthreeia and \texttt{Spacer} cannot solve any of them.
On solved instances in the Q category, \prophic introduced an average of 1.2
prophecy variables and a median of 1. This makes sense because, upon inspection,
most benchmarks only require one quantifier and we are careful to only
introduce prophecy variables when needed.
On benchmarks it cannot solve,
\icthreeia either times out or fails to compute an interpolant. This is expected
because quantifier-free interpolants are not guaranteed over the standard theory
of arrays. Even without arrays, it is also possible for \prophic to fail to compute an interpolant, because MathSAT's interpolation
procedure is incomplete for combinations with non-convex theories such as
integers. However, this was rarely observed in practice.

We further observe that \prophic does not perform as well on unsafe benchmarks.
This is expected, because our array solving procedure is enumeration-based and
should be slower than the array theory solvers within an SMT solver. However, we
believe that a dedicated array solving procedure is important for performance of
the overall algorithm and especially safe benchmarks. We tried minimal
experiments with obtaining array lemmas directly from the SMT solver and did not
achieve comparable performance. This is likely because our array solver is aware
of the ultimate goal to run $\Prophecizenoarg$ with a small delay and can
enumerate array axioms in a corresponding order, starting with index
instantiations that would require the fewest history variables.

There was one discrepancy in our experiments. On
\emph{chc-LIA-lin-arrays\_381} \gspacer disagrees with \quicthree, \texttt{Spacer}, and
\prophic. This is the same discrepancy mentioned in the CHC-COMP 2020
report~\cite{chc-comp-2020-report}. \prophic proved this benchmark safe without
introducing any auxiliary variables and we used both CVC4~\cite{cvc4} and
MathSAT to verify that the solution was indeed an inductive invariant for the
concrete system. We are confident that this benchmark is safe and thus do not
count it as a solved instance for \gspacer.

Some of the tools are sensitive to the encoding. Since it is syntax-guided,
\freqhorn is sensitive to the encoding syntax. The freqhorn benchmarks were
hand-written in CHC to be syntactically simple; this simplicity is maintained by
\texttt{horn2vmt} and also benefits \prophic. However, \prophic can be sensitive
to other encodings. For example, the \textit{quic3} benchmarks translated by
\texttt{SeaHorn} and included in \textit{chc-comp-2020} are much harder for
\prophic to solve (after translation by \texttt{horn2vmt}) compared to the
direct C to VMT translation using \emph{Kratos2}. We found that \prophic solves 6 benchmarks when
translated by \texttt{horn2vmt} $\circ$ \texttt{SeaHorn}, versus 41 when
translated directly by \emph{Kratos2}. We stress that the CHC solvers performed
similarly on both encodings: our experiments showed that \quicthree and
\freqhorn solved exactly the same number in both translations, and \gspacer
solved 27 when translated with \emph{Kratos2} and 34 when translated with
\texttt{SeaHorn}. Importantly, \prophic on the \emph{Kratos2} encoding solved
more benchmarks than any other tool and encoding pair.

There are two main reasons why \prophic fails on the \texttt{SeaHorn} encodings.
First, due to the LLVM-based encoding, some of the \texttt{SeaHorn} translations
have index sets which are insufficient for finding the right prophecy variable.
This has to do with the memory encoding and the way that fresh variables and
guards are used. \texttt{SeaHorn} also splits memories into ranges which is
problematic for our technique. Second, the \texttt{SeaHorn} translation is
optimized for CHC, not for transition systems. For example, it introduces many
new variables, and the argument order between different predicates may not
match. In the transition system, this essentially has the effect of
interchanging the values of variables between each loop. \texttt{SeaHorn} has
options that address some of these issues, and these helped
\prophic solve more benchmarks, but none of these options produce encodings that
work as well as the \emph{Kratos2} encodings. The difference between good CHC
and transition system encodings could also explain the overall difference in
performance on \emph{chc-comp-2020} benchmarks, most of which were translated by
\texttt{SeaHorn}. Both of these issues are practical, not fundamental, and we
believe they can be resolved with additional engineering effort.

\subsection{Self Comparison}
Next, we run a self-comparison using different options in \prophic. We
accomplish this by starting with the configuration used above, and dropping a
single feature to obtain a new configuration. This serves as a metric of how
important each heuristic is to the overall performance of \prophic. Each
configuration has a unique string identifying it as follows:
\begin{enumerate}
  \item \texttt{sa}: with strong abstraction;
  \item \texttt{nav}: no outer CEGAR loop that abstracts large values;
  \item \texttt{na}: no assuming property in pre-state;
  \item \texttt{npr}: no attempting to reduce the number of prophecy
    variables introduced;
  \item \texttt{ntp}: no tracking prophecy variables as important variables
    to guide IC3IA to useful predicates;
  \item \texttt{nur}: no unsat-core based reduction when enumerating timed axioms;
  \item \texttt{nhp}: no seeding IC3IA with predicates obtained from equalities
    between history variables and targets over current state variables, e.g., if
    $h_t^{1\prime} = t$ is in the transition relation, would add $h_x^1 = t$ as
    a predicate;
  \item \texttt{nar}: no additional reduction of consecutive axioms (differs
    from \texttt{nur} in that the consecutive axioms are lifted first);
  \item \texttt{noheur}: a combination of 2-8.
  \end{enumerate}

\iflongversion
\begin{figure}
  \centering
  \includegraphics[width=\linewidth]{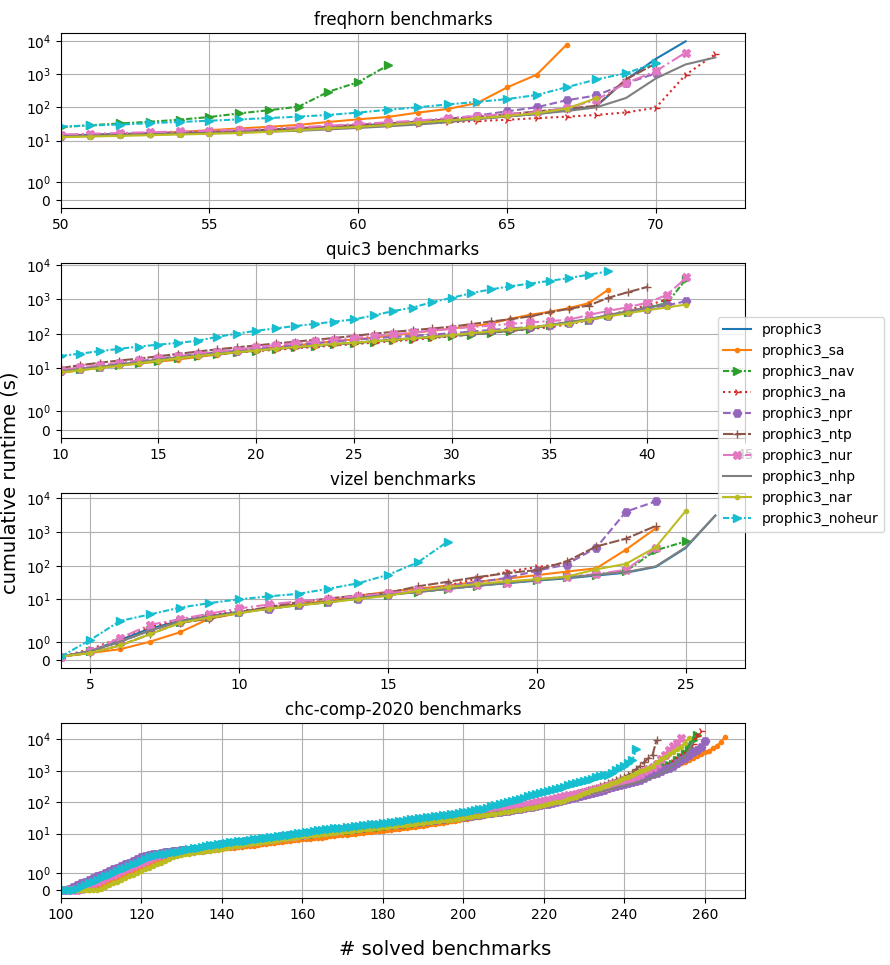}
  \caption{Number of solved benchmarks in self-comparison over time (sorted)}
  \label{fig:cactus-self-comparison}
\end{figure}
\fi

Fig.~\ref{fig:self-comparison} shows the results in a table and
Fig.~\ref{fig:cactus-self-comparison} plots the number of solved benchmarks over
  time. We observe that \prophicSA solves fewer benchmarks in the
  \textit{freqhorn}, \textit{quic3}, and \textit{vizel} sets. However, it is
  faster on commonly solved instances. This makes sense because it needs to
  check fewer axioms (it uses built-in equality and thus does not check equality
  axioms). We suspect that it solves fewer benchmarks in the first three sets
  because it was unable to find the right prophecy variable. For example, for
  the \texttt{standard\_find\_true-unreach-call\_ground} benchmark in the
  \emph{quic3} set, a prophecy variable is needed to find a quantifier-free
  invariant. However, because of the stronger reasoning power of \SA, the system
  can be sufficiently refined without introducing auxiliary variables.
  \icthreeia is then unable to prove the property on the resulting system
  without the prophecy variable, instead timing out. Interestingly, notice that
  \prophicSA solves the most benchmarks in the QF category overall, suggesting
  that there are practical performance benefits of the CEGAR approach even when
  quantified reasoning is not needed.

  Feature \texttt{nav} is the additional CEGAR loop for abstracting large
  values. The results show that this primarily affects the freqhorn benchmarks.
  This is expected because those contained several examples with large, constant
  loop bounds. This means a quantifier was not strictly necessary, but was
  needed in practice. Without abstracting the loop bound,
  Algorithm~\ref{alg:refine-ts} would take far too long to reach spurious
  counterexamples due to the large unrolling bound before an error state is
  reached.

  Based on these results, each of the other heuristics alone do not make a big difference for
  these benchmarks. However, the \texttt{noheur} experiments
  demonstrates that dropping all of them simultaneously does negatively impact
  performance. It is slower overall in the cactus plots of
  Fig.~\ref{fig:cactus-self-comparison}, and solves markedly less in the \emph{vizel}
  and \emph{chc-comp-2020} benchmark sets. The core algorithm performs well
  alone, but the heuristics interact to further improve performance.

\section{Related Work}
\label{sec:related}

We refer often to McMillan's work in~\cite{eager-abstraction}. In that paper,
McMillan reduces infinite-state model checking problems to finite-state problems
that can be checked with a SAT-based model checking algorithm by eagerly
instantiating axioms. Not all possible axioms are instantiated, which is why
this is an eager \emph{abstraction}. This process requires introducing auxiliary
variables. We use several of the same theorems, but for a different goal. Rather
than reducing infinite-state to finite-state systems, we are interested in
reducing problems with quantified inductive invariants to ones with
quantifier-free ones. Furthermore, while the approach of~\cite{eager-abstraction}
is a very general framework that is primarily applied
manually, we focused on infinite arrays and provided a fully automated
algorithm.

There are two important related approaches for abstracting arrays in horn
clauses~\cite{cell-morphing} and memories in hardware~\cite{bjesse}. Both make a
similar observation that arrays can be abstracted by modifying the property to
maintain values at only a finite set of symbolic indices. We differ from the
former by using a refinement loop that automatically adjusts the precision and
targets relevant indices. The latter is also a refinement loop that adjusts
precision, but differs in the domain and the refinement approach, which uses a
multiplexor tree. Although neither paper uses the term \emph{prophecy variable},
their refinement approaches can be viewed as prophecy-variable based. We differ
from both approaches in our use of array axioms to automatically find and add
auxiliary variables.

A similar lazy array axiom instantiation technique is
proposed in~\cite{euficient}. However, their technique utilizes interpolants for
finding violated axioms and cannot infer universally quantified invariants. The
work of~\cite{nra-mc} also uses lazy axiom-based refinement, abstracting
non-linear arithmetic with uninterpreted functions. We differ
in the domain and the use of auxiliary variables. In~\cite{temporal-prophecy},
prophecy variables defined by temporal logic formulas are used for liveness and
temporal proofs, with the primary goal of increasing the power of a temporal
proof system. In contrast, we use prophecy variables here for a different
purpose, and we also find them automatically. The work of~\cite{local-proofs}
includes an approach for synthesizing auxiliary variables for modular
verification of concurrent programs. Our approach differs significantly in the
domain and details.

There is a substantial body of work on automated quantified invariant generation
for arrays using first-order theorem
provers~\cite{vampire,vampire-arrays-workshop,vampire-arrays-orig,saturating-quantified-arrays}.
These include extensions to saturation-based theorem proving to analyze specific
kinds of predicates, and an extension to paramodulation-based theorem proving
to produce universally quantified interpolants. In~\cite{auto-array-invar}, the
authors propose an abstract interpretation approach to synthesize universally
quantified array invariants. Our method also uses abstraction, but in a CEGAR
framework.

Two other notable approaches capable of proving properties over arrays that
require invariants with alternating quantifiers are~\cite{trace-logic,synrg}. The
former proposes \emph{trace logic} for extending first-order theorem provers to
software verification, and the latter takes a \emph{counterexample-guided
  inductive synthesis} approach. Our approach takes a model checking perspective
and differs significantly in the details. While these approaches are more
general, we compared against state-of-the-art tools that focus specifically on
universally quantified invariants.

MCMT~\cite{mcmt,cubicle,cubicle-far} and its derivatives~\cite{safari,booster}
are backward-reachability algorithms for proving properties over ``array-based
systems,'' which are typically used to model parameterized protocols. These
approaches target syntactically restricted \emph{functional} transition systems
with universally quantified properties, whereas our approach targets general
transition systems. Two other approaches for solving parameterized systems
modeled with arrays are~\cite{smt-based-param} and~\cite{i4}. The former
iteratively fixes the number of expected universal quantifiers, then eagerly
instantiates them and encodes the invariant search to nonlinear CHC.\@ The latter
first uses a finite-state model checker to discover an inductive invariant for a
specific parameterization and then applies a heuristic generalization process.
We differ from all these techniques in domain and the use of auxiliary
variables. Due to the limitations explained in Section~\ref{sec:expressiveness}, we
do not expect our approach to work well for parameterized protocol verification
without improvements.

In~\cite{indexed-pred}, heuristics are proposed for finding predicates with
free indices that can be universally quantified in a predicate
abstraction-based inductive invariant search. Our approach is
counterexample-guided and does not utilize predicate abstraction directly
(although IC3IA does). The authors of~\cite{heap-invar} propose a technique for
Java programs that associates heap memory with the program location where it
was allocated and generates CHC verification conditions. This enables the
discovery of invariants over all heap memory allocated at that location, which
implicitly provides quantified invariants. This is similar to our approach in
that it gives quantification power without explicitly using quantifiers and in
that their encoding removes arrays. However, we differ in that we focus on
transition systems and utilize a different paradigm to obtain this implicit
quantification. Prophecy variables have also been proposed for Hoare-style
reasoning about concurrent programs. In~\cite{proph-structural}, the authors
formalize ``structural'' prophecy variables for Hoare logic, which can only
predict state within their own thread. The authors of~\cite{proph-general}
generalize this approach for separation logic to allow predicting values between
different threads. Our work differs in the domain and level of automation.

\section{Conclusion}
\label{sec:conclusion}
We presented a novel approach for model checking transition systems containing
arrays. We observed that history and prophecy variables can be extremely useful
for reducing quantified invariants to quantifier-free invariants. We
demonstrated that an initially weak abstraction in our CEGAR loop can help us to
\emph{automatically} introduce relevant auxiliary variables. Finally, we
evaluated our approach on four sets of interesting array-manipulating
benchmarks. In future work, we hope to improve performance, explore a tighter
integration with the underlying model checker, address the limitations described
in Section~\ref{sec:expressiveness}, and investigate applications of
counterexample-guided prophecy to other theories.

\section*{Acknowledgment}
\noindent This work was supported by the National Science
Foundation Graduate Research Fellowship Program under Grant No.~DGE-1656518. Any
opinions, findings, and conclusions or recommendations expressed in this
material are those of the author(s) and do not necessarily reflect the views of
the National Science
Foundation. 
Additional support was provided by DARPA, under grant No.~FA8650-18-2-7854. We
thank these sponsors for their support. We would also like to thank Alessandro
Cimatti for his invaluable feedback on the initial ideas of this paper.

\iflongversion
\section*{Data Availability Statement}
The experimental results and the necessary software (on the TACAS 2021
Ubuntu 20.04 virtual machine~\cite{tacas21-vm}) for reproducing the results
shown in Figure~\ref{fig:reorganized-results} are available in the
Figshare repository: \url{https://doi.org/10.6084/m9.figshare.13619096}.
\fi

\bibliographystyle{alphaurl}
\bibliography{ms}

\end{document}